\documentclass[a4paper]{jpconf}
\usepackage{graphicx}
\usepackage{epsfig}
\def\be{\begin{equation}}
\def\ee{\end{equation}}
\def\bea{\begin{eqnarray}}
\def\eea{\end{eqnarray}}
\def\aprle{\buildrel < \over {_{\sim}}}
\def\aprge{\buildrel > \over {_{\sim}}}
\begin{document}

\title{Neutrino-2008: Where are we? Where are we going?}

\author{Alexei Yu. Smirnov\footnote{Invited talk 
at the XXIII International Conference on Neutrino Physics and Astrophysics, Christchurch 
New Zealand, May 25 - 31, 2008}
}

\address{International Centre for 
Theoretical Physics, Strada Costiera 11, 34014 Trieste,  Italy,\\ 
Institute for Nuclear Research, RAS, Moscow, Russia}

\ead{smirnov@ictp.trieste.it}

\begin{abstract} 
Our present knowledge of neutrinos 
can be summarized in terms of the ``standard neutrino scenario''. 
Phenomenology of this scenario as well as  attempts to uncover physics 
behind neutrino mass and mixing are described. Goals of future studies 
include complete reconstruction of the neutrino mass and flavor 
spectrum, further test of the standard scenario and  search for 
new physics beyond it.  
Developments of new  experimental techniques may lead to 
construction of new neutrino detectors from table-top   
to multi-Megaton scales which will open new horizons in the field. 
With detection of neutrino bursts from the Galactic supernova and 
high energy cosmic neutrinos neutrino astrophysics will enter 
qualitatively new phase.  
Neutrinos and LHC (and future colliders), neutrino astronomy, 
neutrino structure of the Universe, and probably,  neutrino technologies 
will be among leading topics of research.

\end{abstract}

\section{Introduction}

``Where are we?'' 
\footnote{The author does not take  responsibility for the title of his 
talk. Still, he will do his best to make sense out of it.} 
is a beloved question of neutrino physicists, 
which may be explained by the elusive character of the subject 
 of research. According to the HEP Spires 
from 52 papers with such a title, 13  are on neutrinos. 
John  Bahcall keeps the record: 6 papers 
(all on solar neutrinos; in the field where real progress has been achieved).  
Of course, not only neutrino physicists are lost, however the  
number of papers in other fields is substantially smaller:   
``where we are'' in particle physics, 
in heavy ion collisions, in non-baryonic dark matter, 
in high energy physics have been asked 2 times each. 
String theory? -  1 time ... 

Encouraging:  Glashow, Lederman and Weinberg 
were among those who asked.   
Variations on the theme:   
``How it started and where we are?'' ``Where we are 
and where do we stand?'' ``Where should we go?'' 
``Where are we coming from?''  
and even more profound: ``Who we are?'' (J. Ellis, with reference to 
P. Gauguin).  
 
Let us elaborate further: 
Where are we in time? By the way, the New Zealand time was introduced in 1868 - 
40 years before the year of the  Rutherford's Nobel prize award - our starting point.  
Then 

1928 - Dirac equation.

1938 - Majorana, his disappearance. 

1948 - Gardner and Lattes: artificial production of pions. 

1958 - Goldhaber, Grodzins, Sunyar: Helicity  of neutrino (50th anniversary!).  

1968 - Davis: the first solar neutrino result - the 
birth of the solar neutrino problem. 

1978 - Wolfenstein: ``Neutrino oscillations in matter''.

1988 - Kamiokande-II: the birth of the atmospheric neutrino 
problem \footnote{L. Sulak has informed me on some earlier indications 
of the anomaly in the IMB results.} .

1998 - Discovery of oscillations in atmospheric neutrinos. 

2008 - Discovery of New Zealand by the neutrino community; 
the start of LHC.  

Where are we in space? - about 5000 km (baseline) from IceCube in 3D, 
somewhere in the electroweak brane, in extra D... 

Where are we in the field of neutrino physics? 
The answer includes:   
``conquest territory'' - the standard neutrino scenario (sec. 2); 
understanding neutrino masses and mixing (sec. 3);
beyond the standard scenario (sec 4);  
a future which we know (sec. 5); a future which we can only imagine (sec. 
6).

\section{Standard neutrino scenario}

\subsection{Standard scenario}  
``Standard neutrino scenario'' 
can be formulated in the following way: 

\begin{itemize}
\item
Neutrino interactions are described by the standard electroweak model.

\item
There are only 3 types of light neutrinos (three flavor and three mass states).  

\item 
Neutrinos are massive. Neutrino masses  
   are in the sub-eV range - much smaller than 
   masses of charged leptons and quarks. 

\item
Neutrinos mix. There are two large mixing angles  
and one small or zero angle. The pattern of lepton 
    mixing  strongly differs from that of quarks.

\item
The observed masses and mixing have pure vacuum origin;
they are generated at the electroweak, and probably,  higher energy scales. 
These are ``hard'' masses. 

\end{itemize}

The standard scenario is a result of work of several generations 
of neutrino physicists, 
the collective effort of experimentalists and theoreticians \cite{appol}. 
This scenario is our ``conquest territory'', basis and starting  point 
for further advance,  summary of results of the first phase of studies 
of neutrino mass and mixing. 
The following comments are in order. 

1). Interactions:  The gauge interactions of neutrinos are well known  
and well checked. In contrast, there is no information about 
the  Yukawa couplings with the Higgs boson    
(if the RH neutrinos exist); these couplings  
can be  relevant for leptogenesis. 
Neutrino interactions with complex systems: nucleons 
and nuclei are not completely understood and  
open questions are related to the physics of strong interactions. 
As a probe, neutrinos are unique, being sources of the axial vector currents. 
The open questions include the value of axial 
mass in the quasi-elastic scatterings 
\cite{axial}, the coherence in a  single pion forward production 
\cite{singlepi}, the role of the axial vector anomaly  
in interactions of $Z$, $\gamma$, $\omega$  
in explanation  of  the low energy excess 
observed in the MiniBooNE experiment \cite{hill}. 
Significant progress has been achieved 
in nuclear physics for $\beta\beta$-decays   
\cite{bbonu}.
Rare neutrino processes and processes at extreme conditions relevant 
for astrophysics, e.g., $\nu \bar{\nu}-$ pair production in 
nucleon collisions, are  under consideration.  

2).  Propagation:
There are still some discussions about the  theory 
of neutrino oscillations even in vacuum. 
``Eternal questions'' include  the equality of momenta or energies in 
consideration of interference, validity and applications of the 
stationary source approximation,  
relevance of the wave packets, coherence, role of recoil 
of accompanying  particles, {\it  etc.}. 
Some of these issues 
have just an academic interest in normal situation,  
but become important for oscillations at extreme conditions, {\it  e.g.}, 
oscillations of ``Moessbauer neutrinos'' \cite{moessbauer},  
where the uncertainty in energy is much smaller than the oscillation  
frequency: $\Delta E \ll \Delta m^2/2E$ \cite{moess-osc}. 

Concerning propagation in a medium, the forefront of studies has shifted 
to extreme conditions - high densities, 
temperatures, magnetic fields, propagation in neutrino gases, {\it etc}..
Collective non-linear effects induced by the  $\nu\nu-$ scattering 
are in explorative phase and serious progress 
has been achieved since 1993 \cite{collect} - \cite{Duan:2007bt}. 

3). Mass and mixing. 
The main line of thinking is that the right handed components of neutrinos  
exist, neutrinos are the Majorana particles, seesaw 
is realized and  the smallness of mass is   
due the existence of some high mass scales (large masses of RH neutrinos).    
The difference of the quark and lepton mass spectra and mixing 
patterns is related somehow to the smallness of the neutrino mass. 
In general, 
\be
m_\nu = m_{hard} + m_{soft} (E, n), 
\ee
where $m_{soft} (E, n)$ is the medium-dependent soft 
component which can be substantial for neutrinos but not for other particles. 
An important phenomenological and experimental 
problem is to put model independent limits on (or discover?)
$m_{soft}$.   

\subsection{Phenomenology.}


To a large extend phenomenology of the standard scenario has 
been elaborated, in some cases - in great details. 
Still some areas exist (cosmic, supernova neutrinos) 
where active research continues now. 
Few spots have not been covered yet. 

1). Solar neutrinos. 
A complete description of physics of conversion  has been elaborated and 
very precise analytic results  have been obtained.  
From experimental side some points are still missing. These include   
detection of 

- the Earth matter effect: day-night asymmetry, zenith angle 
dependence of the signal; 

- upturn of the energy spectrum of boron neutrinos at 
low energies (see, however, the recent BOREXINO result \cite{BOREXINO}) ;

- neutrinos in the so called  ``vacuum-to-matter'' transition  
region (N, O, pep);   

- the pp-neutrinos, and - on the other side -  the hep-neutrinos.

These measurements will provide 
further tests of the LMA solution and matter effects 
in general, they  will open additional  possibilities to  search 
for new physics.  The measurements may shed some light on various  
astrophysical issues, e.g.,  the role of 
CNO cycle, abundance of the heavy elements 
at the surface of the Sun and initial conditions 
for solar evolution \cite{astroprob}, {\it etc.}. 

\begin{figure}[htb]
\centering
\includegraphics[width=90mm]{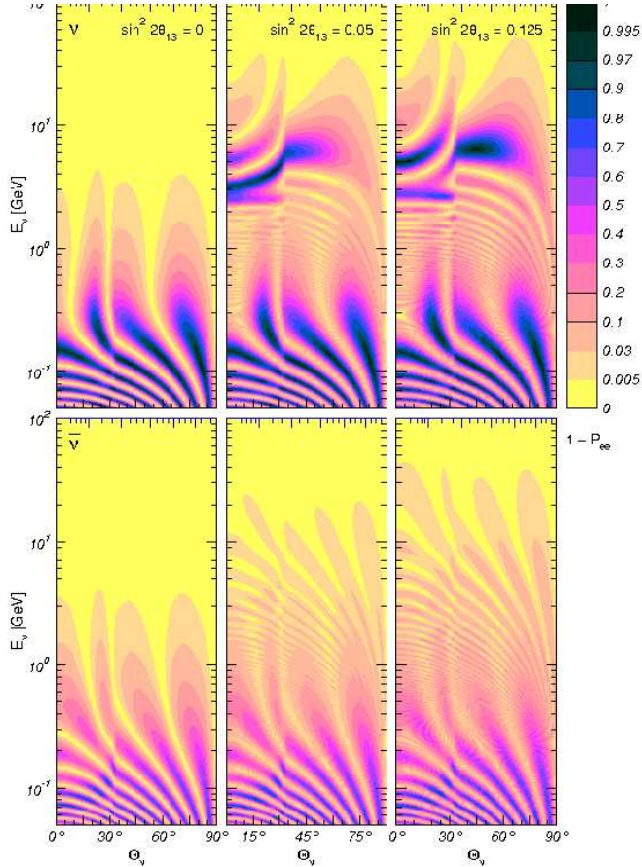}
\begin{minipage}[b]{14pc}
\caption{Neutrino oscillograms in the $3\nu$-mixing case. Shown are the
    contours of constant probability $1 - P_{ee}$ (upper panels) and
    $1 - P_{\bar{e}\bar{e}}$ (lower panels) for $\Delta m^2_{21} = 8 \times
    10^{-5}$~eV$^2$, $\tan^2\theta_{12} = 0.45$ and three different values of
    $\theta_{13}$.   Here $P_{ee}$  is the $\nu_e - \nu_e$ survival probability.  
Normal mass hierarchy is assumed. 
Up to small interference effect the inverted mass hierarchy 
would correspond to interchange of the upper and lower panels. From ref. \cite{AMS2}. 
 \label{fig:solar}}
\end{minipage}
\end{figure} 

2). Atmospheric neutrinos.   
Comprehensive description of neutrino propagation through the Earth
is given in terms of neutrino oscillograms of the Earth -  
lines of equal probabilities 
in the neutrino energy - nadir angle, $E - \Theta_\nu$, plane,  fig. \ref{fig:solar}.  
The oscillograms give a global view on the oscillation phenomena 
inside the Earth  being relevant also for the accelerator 
and cosmic neutrinos\cite{AMS2}.  
The oscillograms are 
the neutrino images of the Earth. 
The Earth is unique and the structures of oscillograms 
seen in fig.~\ref{fig:solar}  are unique and well defined.  
They are defined by the generalized amplitude and phase conditions.  
The former is reduced to the MSW resonance condition 
for one layer (mantle) and to the parametric resonance condition
for 3 layers  (mantle crossing trajectories). 

\begin{figure}[htb]
\centering
\includegraphics[width=80mm]{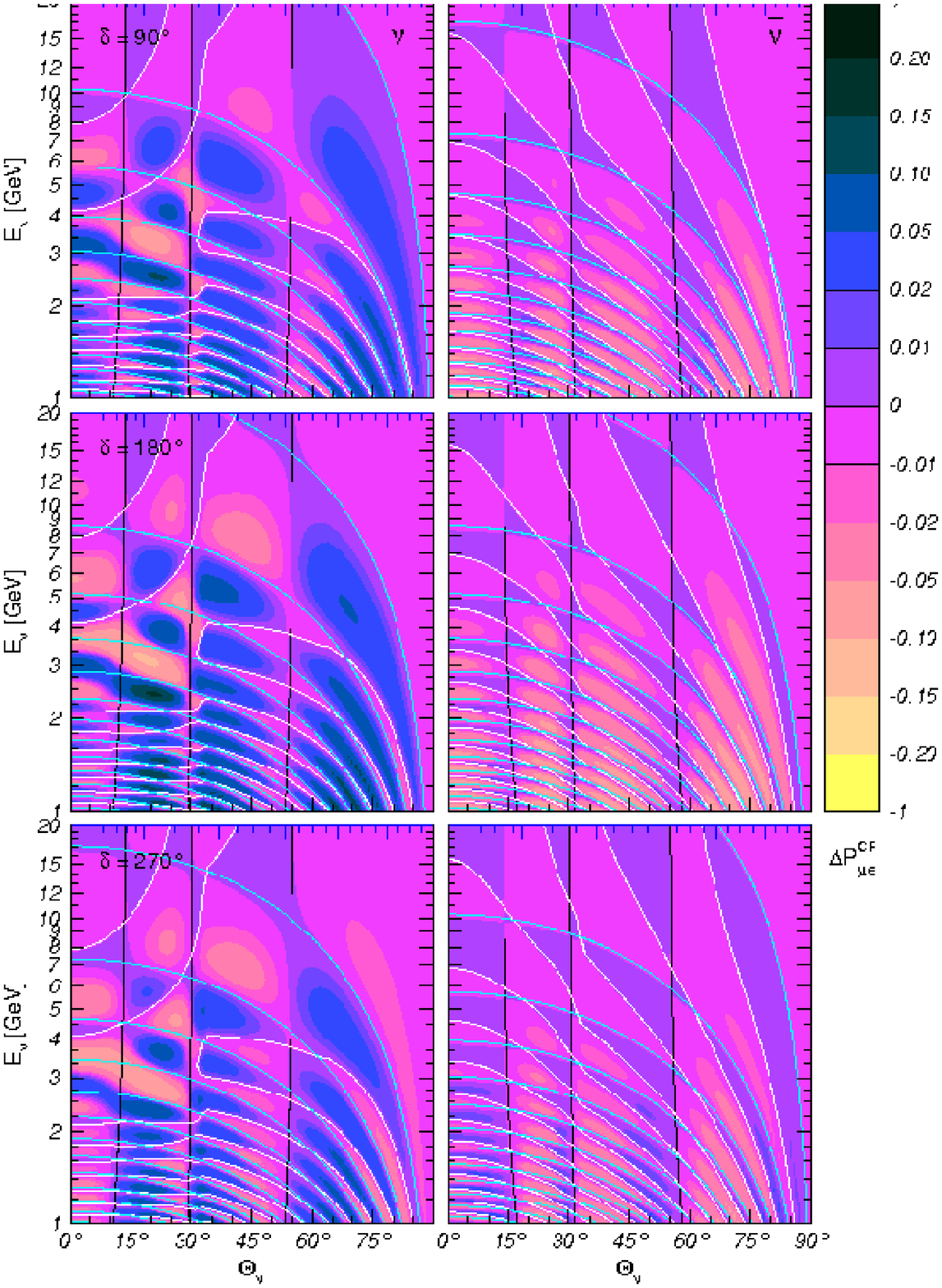}
\begin{minipage}[b]{14pc}
\caption{
    Oscillograms for the difference of probabilities 
$\Delta P_{\mu e}^{CP}(\delta) = P_{\mu e}(\delta) - P_{\mu e}(\delta_0)$
    with $\delta_0 = 0^\circ$. Shown are the solar (black),
    atmospheric (white) and interference phase condition (cyan)
    lines. The lines form the borders of the CP-domains.  
Non-coincidence of the lines and contours 
of $\Delta P_{\mu e}^{CP}(\delta) = 0$ 
from numerical computations is mainly due to the level crossing phenomenon.
The normal mass hierarchy and $\sin^2 2\theta_{13} = 0.05$ are assumed.  
From ref. \cite{AMS2}.
\label{me} }
\end{minipage}
\end{figure}

The CP-violation properties of the oscillograms  
have  the domain structure (see fig.~\ref{me} for the $\nu_\mu \to \nu_e$ channel). 
The $\delta$-dependence appears via the interference term: 
$P_{\mu e}^{int} \sim |A_{A} A_{S}| \cos(\phi - \delta)$,  
where $A_{A}$ and  $A_{S}$ are (in the first approximation) 
the ``atmospheric'' and ``solar'' $2\nu-$amplitudes 
correspondingly and $\phi \equiv arg(A_{S}^* A_{A})$
is the interference phase. 
To assess the $\delta$-dependent terms, one can  consider 
the difference of the oscillation probabilities for two different
values of the CP-phase:
$\Delta P_{\mu e}^{CP}(\delta) \equiv P_{\mu e}(\delta) - P_{\mu e}(\delta_0)$. 
The equality  $\Delta P_{\mu e}^{CP} = 0$ 
holds, if at least one of the following three conditions is fulfilled
\be
\label{eq:abc}    
A_{S}(E_\nu, \Theta_\nu) = 0, ~~~
    A_{A}(E_\nu, \Theta_\nu)  =  0, ~~~ 
    \phi(E_\nu, \Theta_\nu) = (\delta + \delta_0) / 2 + \pi l \,.       
\ee
These equalities determine the  solar and  atmospheric ``magic''  
lines and the interference phases  lines \cite{AMS2} in fig.~\ref{me}  
along which the CP-violation effects are zero.  These lines 
give the borders of the CP-violation domains, the CP-violation has different 
sign in the neighboring domains and 
strong CP-violation is in their central parts.

3).  Long baseline experiments. 
The physics is well understood (see the oscillograms). 
A number of analytic and semianalytic results have been obtained 
and approximate expressions for probabilities were derived 
which use various expansions (perturbation theories) in specific ranges of 
energies and baselines \cite{lblrev}. 
We can speak about the LBL industry:
numerical codes  have been developed which allow one to 
determine sensitivities of experiments 
to unknown parameters:  e.g. $\theta_{13}$ or 
phase $\delta$ using characteristics of experiment (neutrino energy, baseline, 
experimental uncertainties, etc.) as input parameters 
\cite{globes}.

\begin{figure}[h]
\begin{center}
\includegraphics[width=13pc]{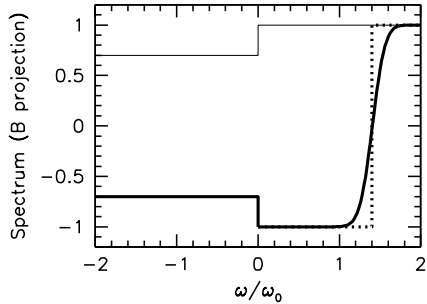}\hspace{2pc}%
\begin{minipage}[b]{13pc}
\caption{\label{boxsp}Neutrino spectra 
for an initial box 
spectrum with 70\%  antineutrinos and initial mixing angle 
$\sin2\theta_{\rm 
eff} = 0.05$.
Negative frequencies correspond
to antineutrinos.  Thin line: initial. Thick dotted: final adiabatic.
Thick solid: numerical example.}
\end{minipage}
\end{center}
\end{figure}

\begin{figure}[h]
 \begin{center}
\includegraphics[width=85mm]{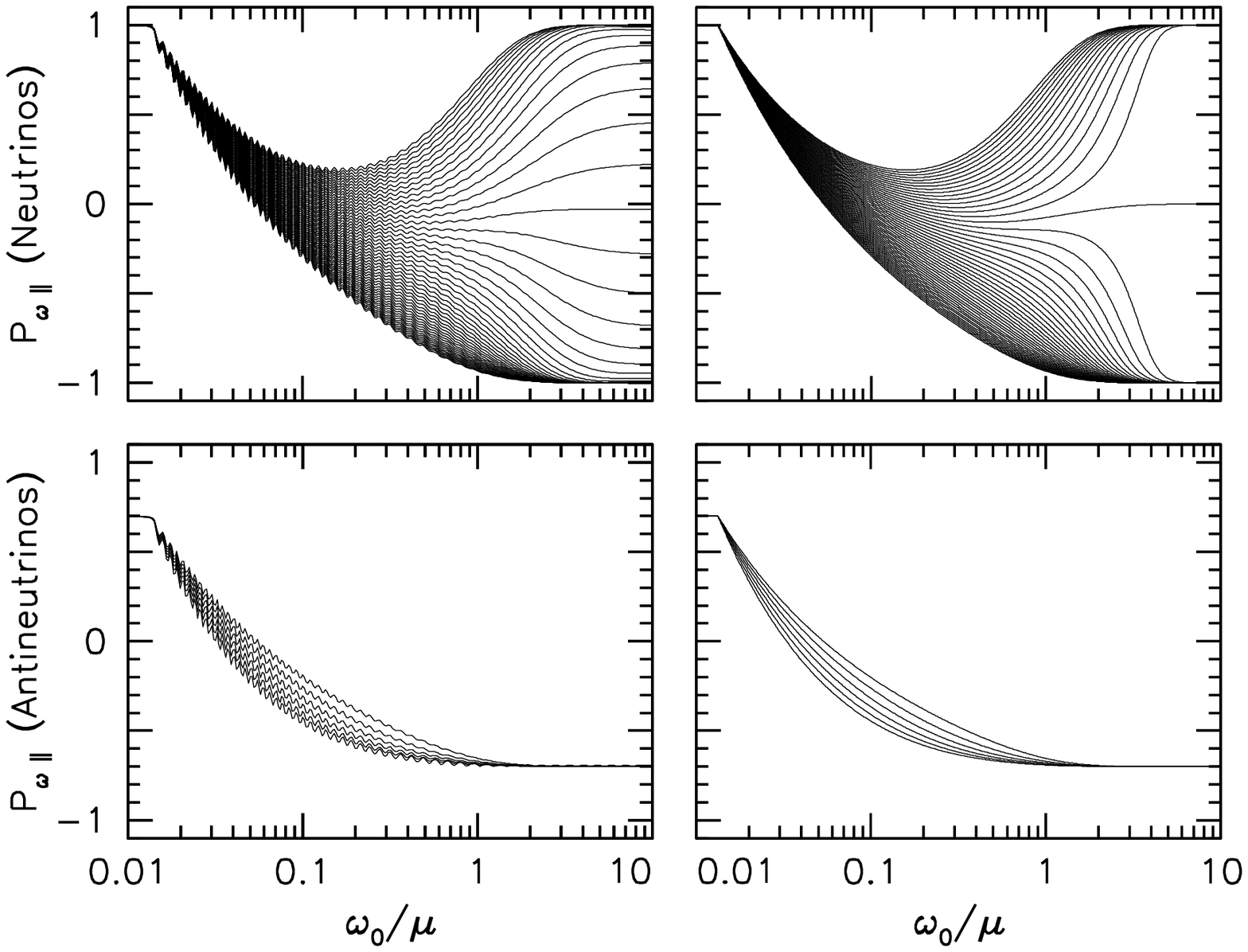}
\begin{minipage}[b]{13pc}
\caption{\label{fig:nunubarspectrum}
$P_{\omega B}(\mu)$ for individual modes for the case of
neutrinos plus antineutrinos. {\it Left:} Numerical solution; 
for $\sin2\theta_{\rm eff} =0.05$.
 {\it Right:} Adiabatic solution for $\sin2\theta_{\rm eff} = 0$. In each
case neutrinos with 51 modes (top) and antineutrinos with 6 modes
(bottom).  From ref.\cite{split}. 
}
\end{minipage}
\end{center}
\end{figure}

4).  Supernova neutrinos.
The $\nu\nu-$ scattering leads to the  flavor exchange 
and variety of collective (non-linear) effects \cite{collect} - 
\cite{Duan:2007bt}. 
One of them  uncovered recently 
is the spectral split \cite{fullersplit,split} 
or swap according to terminology in \cite{Duan:2007bt}.  
An example of the split in a system 
of neutrinos and antineutrinos is shown in figs.~\ref{boxsp},  
\ref{fig:nunubarspectrum}.  
In fig.~\ref{fig:nunubarspectrum}  from \cite{split} the evolutions of 
the ${\bf B}$--components  (or projection on the mass 
axis) of the polarization vectors are shown as functions of 
strength of the $\nu\nu-$ interactions $\mu \equiv \sqrt{2} G_F n_{\nu}$, where
``up'' ($+1$ projection) corresponds approximately to the $e$--flavor and ``down'' 
($-1$)- to the $x$--flavor. 
According to the figure, all modes with frequencies  
below the split frequency: $\omega \equiv \Delta m^2/2E   < \omega_{\rm split}$,  
change  
flavor, whereas the ones with $\omega > \omega_{\rm split}$  
first evolve in flavor space  but then return to their original flavor. 
The split may lead to observable consequences. 
The relevance of these effects for real supernovas still should be clarified.  


4). Cosmic neutrinos.
This is the field of active studies which moves now 
to qualitatively new level.
A number of developments is related to 
recent results in the  $\gamma-$ astronomy 
($\nu - \gamma$   connection,  
implications of the EM radiation data).   
The developments were also triggered by forthcoming large 
scale experiments (IceCube, ANTARES). 
Among possible sources of neutrinos are AGN, GRB, 
core collapse supernovae, SN remnants, microquasars, blasars  
\cite{cnusources}. 
Detailed computations of the neutrino yield  
have been performed for  different conditions in the sources.  
Various effects of neutrino propagation are under consideration: 
vacuum oscillations, conversion in matter 
of the source,  effects of non-standard interactions. 
For maximal 2-3 mixing the original flavor ratio equals  
$F_e, F_\mu, F_\tau \approx 2:1:0 $, which is realized in the case of 
free decays $\pi \rightarrow \mu \nu_{\mu}\rightarrow 
e 2\nu_\mu \nu_e$. Oscillations ``equilibrate'' flavors 
and  the ratio becomes $1 : 1 : 1$. 
Measurements of the ratio and searches for deviations from 
equilibration will be one of the main goals of neutrino astronomy \cite{cnuflratio}. 
The deviation can be due to matter effects in the source,  
various non-standard interactions,  
deviation of 2-3 mixing from the maximal one, contributions 
from other possible mechanisms of neutrino production.  
One of interesting possibilities is neutrino from thick sources: 
Protons are accelerated in the relativistic jets 
by the inner shocks and neutrinos are produced in 
the $pp-$ and $p\gamma-$ collisions. Flavor conversion 
occurs in the He- and H- envelopes \cite{razz}. It leads to breaking of 
flavor democracy. There are new recent developments 
related  to establishing the GZK cut-off and  evidences that  AGN 
are the sources of the  cosmic rays. 
Perspectives to see the cosmogenic neutrinos 
will be further clarified.


\section{Where are we in understanding the neutrino mass and mixing?}

\subsection{Theory of neutrino mass}
In recent years there was an enormous theoretical activity 
in attempt to understand origins of neutrino mass and mixing, 
to explain the smallness of neutrino mass and 
peculiar mixing pattern.  
The simplest possibilities have been explored. A number of 
approaches and scenarios of physics beyond the standard model 
were proposed. Clearly, with only one theoretical talk \cite{sking} 
the program of the conference does not reflect this activity. 
Reason? Nothing is really accomplished? No progress? 
Recall, we measure all these $\theta_{13}$, $\delta$, {\it etc.},   
to uncover eventually the underlying physics, 
to make on this basis new testable predictions.  
Another aspect of the measurements is neutrino applications. 
The whole excitement was that neutrino mass and mixing  
are manifestations of physics beyond the standard model.
Dramatically, after many years of studies and many trials 
the underlying physics has not been identified. 
We should  explore how the progress can be achieved.

\subsection{Bottom-up}
There are three lines of studies in the bottom-up approach  
with different implications for fundamental physics 
and different connections between leptons and quarks.

1). Tri-bimaximal mixing (TBM).  
Immediate implication: flavor symmetry.  
The majority of models proposed so far are based on the 
discrete symmetry group $A_4$.  
Other possibilities explored in this connection  
include  the groups 
$T'$, $D_4$, $S_3$, $S_4$, $\Delta (3n^2)$. 
Extension of these symmetries to quarks is, however, problematic, it requires 
further complication of models.  
TBM may indicate that quarks and leptons are  fundamentally different. 
Mixing and masses are not related  at least in a straightforward way.  

2). Quark-Lepton Complementarity (QLC) 
is based on observations that 
$\theta_{12}^l + \theta_{12}^q \approx \pi/4$ and 
$\theta_{23}^l + \theta_{23}^q \approx \pi/4$. 
A general scheme is  ``the lepton mixing = bi-maximal mixing - CKM''.   
Two extreme realizations of the complementarity, $QLC_\nu$  and  $QLC_l$,  are 
determined  by the order of the bi-maximal and CKM rotations: 
\be
U_{PMNS} =  U_{bm} U_{CKM}^{\dagger}~~(QLC_l), ~~~~ 
U_{PMNS} = U_{CKM}^{\dagger} U_{bm}, ~~  (QLC_{\nu}). 
\label{qlcnu}
\ee 
Implications: Quark-lepton symmetry, or grand unification (GUT), 
plus the existence of structure which produces the bi-maximal mixing. 
The latter may require some symmetry.  
Again there is no straightforward connection between  mixing and masses.

3). Quark-lepton universality. This  approach  does not rely on any specific 
symmetry in the lepton sector. 
The mass (Yukawa coupling) matrices of quarks and leptons  
have no fundamental distinction. The whole difference 
is related to the seesaw mechanism itself  
which explains simultaneously the smallness of neutrino mass and  
large lepton mixing. The mass matrices of quarks and leptons are constructed 
on the basis of the same principles 
(e.g. Froggatt-Nielsen mechanism,  $U(1)-$ flavor symmetry), 
and furthermore,  masses and mixing are related with each other. 
Large lepton mixing can be associated to the  
weak mass hierarchy of neutrinos.

TBM and two versions of QLC differ by predictions 
of the mixing angles $(\theta_{12}, \theta_{13})$: 
\be
QLC_{\nu}: ~ (35.4^{\circ}, 9^{\circ}),   ~~~
TBM:~  (35.2^{\circ},  0),   ~~~
QLC_{l}: ~ (32.2^{\circ}, 1.5^{\circ}). 
\ee 
Notice that $\theta_{12}(QLC_l)  = \pi/4 - \theta_C$  and 
$\theta_{12}(QLC_{\nu})  \approx  \theta_{12}(TBM)$. 
All three  possibilities (subject to RGE corrections) 
agree with the present data within $1\sigma$. 
Clearly, a combination of future precise measurements 
of these angles will  disentangle the schemes.   
In specific models, some additional 
corrections appear due to violation of the underlying symmetry. 
Small deviations from the predictions do not exclude the context. 
Exact confirmation would be very demanding and restrictive. 

There is no reason to consider  TBM but ignore the  
Koide relations which are, in contrast to TBM,  the pure mass relations \cite{koide}.  
Furthermore, it may happen that some  connection between the Koide relation and TBM 
exists. Recall, the equality 
\be
\frac{m_e + m_\mu + m_\tau}
{(\sqrt{m_e} + \sqrt{m_\mu} + \sqrt{m_\tau})^2} = 
\frac{2}{3} 
\label{koide1}
\ee
is satisfied with accuracy $10^{-5}$ on the mass shell and with  $10^{-3}$ -  
at $M_z$. 
The equality (\ref{koide1}) 
has been obtained in attempts to explain relation between the Cabibbo angle 
and lepton masses. 
Both relations can be reproduced if 
\be
m_i = m_0 (z_i + z_0)^2, ~~~ \sum_i z_i = 0,~~~ 
z_0 = \sqrt{\sum_i z_i^2/3},  
\ee
where $z_i$ are some numbers. 
Brannen \cite{brannen} has  generalized the relation to neutrinos: 
\be
\frac{m_1 + m_2 + m_3}
{(- \sqrt{m_1} + \sqrt{m_2} + \sqrt{m_3})^2} = 
\frac{2}{3},  
\label{brannen}
\ee
where the minus sign in front of the first term in denominator is crucial. 
According to (\ref{brannen}) neutrinos have a  
hierarchical spectrum with $m_1 = 3.9 \cdot 10^{-4}$ eV. 
 Non-abelian flavor symmetry and  specific VEV alignment can be 
behind the relations.

\subsection{Flavor symmetries}  
Flavor features of various symmetry groups have been explored: 
Discrete groups $A_4$  (subgroup of $SO_3$) 
and $T_7$ -  Frobenius group (subgroup of $SU_3$) \cite{t7}
look rather promising. 
It was argued that the minimal group which 
leads to TBM mixing is $S_4$ \cite{lam}. 
The ``successful'' models imply tuning 
of symmetries and patterns of their breaking. The following aspects 
are of special interest. 

1).  Fundamental {\it versus} effective. 
The required symmetry may appear only  at the 
{\it effective} level after decoupling  of heavy degrees of freedom.
No flavor symmetry or some other symmetry exist at the 
fundamental level.  
In the case of decoupling of the RH neutrinos 
the emerging symmetry can be called the  ``see-saw symmetry'' \cite{bilash}. 
This idea is along with the line of Ref. \cite{ferretti}, 
where it was argued that symmetries at the effective 
level may follow from certain hierarchies of masses at the fundamental level.

2). Real {\it versus} accidental? 
Are the observed flavor features, such as maximal 2-3 mixing, tri-bimaximal mixing, 
small (zero) 1-3 mixing, Koide relations accidental? 
Some value of mixing angle  is accidental 
if it is a combination  of two or more independent contributions.  
If some value or relation appears as immediate ``one-step'' 
consequence of symmetry (the group structure),   we conclude that they are 
not accidental, that is, a  real.  The decisive criteria 
are new testable predictions from symmetries. 
Discovery of the degenerate mass spectrum would be 
convincing evidence of a real  symmetry. 

3). Flavors and GUT. 
The scale of  RH neutrino masses favors of GUT. 
In fact, the value of mass of the heaviest RH neutrino can coincide with 
the GUT scale $ M_R \approx M_{GUT} \sim 10^{16}$ GeV, 
which can be achieved in the presence of mixing of three generations. 
Alternatively, the scale of RH neutrino masses can be related 
to $M_{GUT}$ via the Planck scale $M_{Pl}$: 
$M_R \approx M_{GUT}^2/M_{Pl} \sim 10^{14}$ GeV  
(the latter is realized, e.g., in the double seesaw scenario).
Another indication of GUT is  QLC. 
The generic problem of unification of quarks and leptons is the difference 
of their  mixing patterns.  
To explain data with flavor symmetries,  
the quarks and leptons, the RH components of charged leptons and 
neutrinos should have different flavor properties 
This prevents their unification,  
or the original flavor symmetry should be broken differently in 
quark and lepton sectors,  for up and down components of multiplets.  


The data on masses and mixing show both order (regularities) and some 
degree of randomness, and no simple parametrization is found.  
Therefore no simple ``one-step'' explanation 
is expected. Furthermore, different pieces of data testify for 
different underlying physics. This may indicate that several unrelated  
contributions to the neutrino mass matrix exist (zero order structure  plus small 
corrections?). Keeping this in mind one can develop the following 
approach: (i) refrain from attempts to explain all the data at once;  
(ii) take the most symmetric and minimal context ``GUT  plus  
flavor symmetry'', (iii)  explore how far one can go in  
explanation of the data. One possibility is   
$SO(10)$,  without  {\bf 126} Higgses but with non-renormalizable 
operators, with  flavons and singlet fermions. Flavons and singlet fermions 
(their number can be bigger than three) can compose a  
hidden sector of theory with certain symmetries and dynamics.   
In this context one can disentangle the hierarchies 
of quark and neutrino masses and, e.g., relate the  
geometrical hierarchy of the up quark masses and 
nearly maximal 2-3 leptonic mixing \cite{hagedorn}. 

4). Energy scales of new physics. 
In the ``seesaw approach'' the smallness of  neutrino mass 
is in general related to the existence of some new large scale, $\Lambda$. 
In the simplest version $\Lambda$ is just the bare mass of the RH neutrino. 
In general, there is some particle sector and dynamics behind. 
Various realizations have been proposed  with 
$\Lambda$  equal  $M_{Pl}$ (which requires many RH neutrinos), or 
$M_{GUT}$, or $\sqrt{M_{Pl} M_{EW}}$, or $M_{EW}$. 
In $\nu MSM$ scenario \cite{numsm} $\Lambda  < 0.1 - 0.5$ GeV. Even 
the  extreme possibility, $\Lambda = few$ eV, is not excluded \cite{degouvea}.  
All this means that ``Physics behind the neutrino mass'' is not yet identified.

\section{Beyond the standard scenario} 

There are two aspects of further experimental and phenomenological studies: 
tests of the standard scenario and searches for new physics. 
The ways new physics appears in our considerations 
can be classified as follows: 
1). Neutrino anomalies. 
Recall that neutrino anomalies were the driving force of 
the developments for more than 40 years. 
2). New physics related to explanation of the neutrino masses. 
3). New physics motivated by other fields. This includes 
various extensions of the standard model: Left-Right symmetric models, supersymmetry, 
GUT, extra dimensions. 
4). Unmotivated (explicitly) speculations.

\subsection{Neutrino anomalies}
Neutrino anomalies can be considered as potential seeds of new 
developments.  The anomalies up to date 
are summarized in the Table 1.

\begin{center}
\begin{table}[h]
\caption{\label{osc-tab}
Neutrino anomalies}
\centering
\begin{tabular*}{\columnwidth}{@{\extracolsep{\fill}}@{}lll@{}}
\br
Name:  & Feature & possible interpretations\\    
\mr
LSND  & excess of $e^+$ events  & $(exotics)^2$, see text \\
MiniBooNE & excess of events at $E< 400$ MeV  & see text\\
NuTeV  &  value of $\sin \theta_W$  &  structure functions,\\
       &                            &     new heavy leptons \\
Homestake & low rate, tension with other data &  mixing with very light \\ 
     ~ &   ~  &                         sterile neutrino; \\
 ~ &   ~  &               unparticle physics \\
 
Gallium &  deficit of observed signal  & cross-section;  \\
~ & in calibration experiments  & small scale oscillations\\
Unnamed & time variations of solar neutrino signals & 
                             neutrino magnetic moment, \\
\, &   \,  &             periodicity of the energy release \\ 

SN1987A     & angular, time distribution, LSD signal & 
                                           astrophysics?  \\     
$Z^0$-width   &    $N_\nu^{eff} < 3$  & new heavy leptons \\
GSI  & modulation of exponential decay & systematics ?\\
 \, &   \,  &                 unrelated to neutrinos (?) \\
\br
\end{tabular*}
\end{table} 
\end{center}

LSND after MiniBooNE \cite{MiniB-osc}, and  MiniBooNE after LSND. 
LSND can be reconciled with MiniBooNE in model with 
two sterile neutrinos and CP-violation \cite{malt-sch}. 
Here the problem still exists with short baseline experiments and astrophysics. 
After MiniBooNE an explanation of LSND requires 
even more speculative schemes, so to say, 
$(exotics)^2$: e.g., sterile neutrinos and extra dimensions 
\cite{shortcut}, sterile neutrinos with energy dependent masses, \cite{stermass}, CPT- 
violation and sterile neutrinos \cite{cpt-bar}, 
3 sterile neutrinos and light vector boson \cite{nelson}, 
soft decoherence \cite{farzan}. 
In the last case both ``decoherence'' and ``soft'' are exotic. 
Do we deal here with something  
unusual, not connected to known physics processes? 
An interesting task is to reconstruct from the data  $L-$ and 
$E-$ dependence of the underlying  effect 
in model independent way and check the consistency with the other data.

\subsection{New physics} 
Broadly it can be classified as 
(i) non-standard interactions (NSI);
(ii) new neutrino states, (iii) new  dynamics. 

\begin{figure}[b]
\centering
\includegraphics[width=80mm]{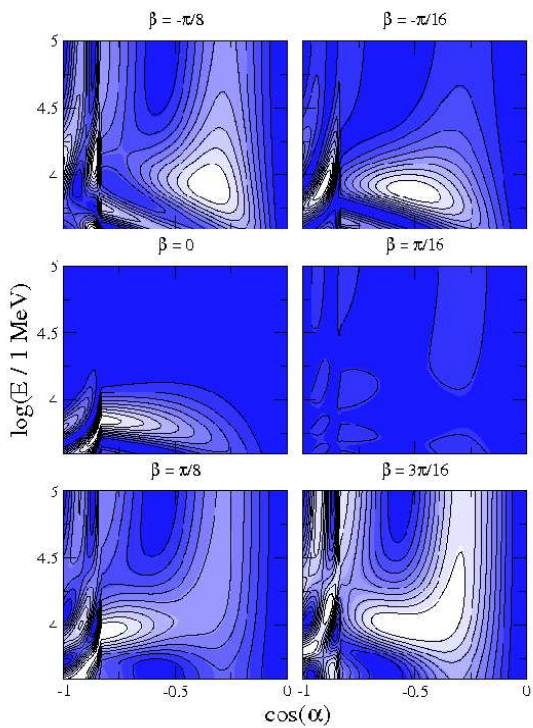}
\begin{minipage}[b]{14pc}
\caption{Neutrino oscillograms of the Earth in the presence of NSI. 
The electron neutrino survival probability, $P_{ee}$, as function of 
zenith angle $\alpha$ and energy. 
Different panels correspond to different strength of NSI,    
$\sin 2\beta \approx - 2 \epsilon_{e\tau}$, $\theta_{13} = 8^{\circ}$.    
Panel with $\beta = 0$ corresponds to the 
the standard interactions only. 
Strong transitions are in white regions. For $\beta = - \pi/8$ and 
$\beta = - 3\pi/16$ these regions extend to high energies.   
From ref. \cite{BO}.
\label{fig:nsi}}
\end{minipage}
\end{figure}

(i) Possible existence of non-standard neutrino interactions  
is related to extensions of SM  at the EW scale and terascale   
as well as  
to new  particles motivated by astrophysics. 
NSI have rich phenomenology influencing both 
propagation and detections of neutrinos. 
In particular, they can modify the  refraction phenomena,
especially at high energies where usual mixing is suppressed,  
see fig.~\ref{fig:nsi} from \cite{BO}, 
 
(ii) New neutrino states or sterile neutrinos. 
If light, these states can have direct observable consequences: 
be produced in various neutrino processes, 
participate in oscillations, and decays, {\it etc.}. 
Mixing of new states with usual neutrinos leads to  indirect effects:  
modifications of the mass matrix of active neutrinos 
(induced mass: $m_{ind} \approx m_S \sin^2\theta_S$), 
breaking of universality, appearance of FCNC.  
Light sterile neutrinos both participate 
in low energy phenomenology and modify the mass matrix 
of active neutrinos. The heavy ones produce indirect effects only. 

For $m_S \aprle  200$ MeV the strong bounds on the mixing 
of sterile neutrinos from astrophysics and cosmology 
exclude significant influence on the mass and mixing of 
active neutrinos. 
In contrast, for $m_S \aprge 1$ GeV, the indirect effects dominate. 
The induced mass terms can generate 
the dominant elements of active neutrino mass matrix.  

3). New dynamics: this includes violation of fundamental symmetries and principles,  
such as CPT, Lorentz invariance, Pauli principle, as well as  
non-standard decoherence,  effects of unparticle physics, etc..  
According to the unparticle physics scenario \cite{unpart}  
the hidden sector (HS) of theory exists which includes 
the gauge theory with fermions.  The number of fermions 
in the HS is  such that the effective gauge coupling $g$
increases with the decrease of energy and at the energies below  
certain scale $\Lambda_U$ approaches 
the infrared fixed point, $g \rightarrow g^*$. 
If $g^* \gg 1$ the fermions form composite (confined) states. 
To some extend this transition is similar   
to the transition from  quarks to hadrons below $\sim \Lambda_{QCD}$. 

The particles of HS couple to the SM particles via the exchange of 
messenger fields with mass $M \gg \Lambda_U$.
At energies below $M$ the interaction 
of SM particles with HS particles are described 
by the effective interactions  
$\frac{1}{M^k} O_{SM} O_{UV}$, 
where $O_{SM}$ and  $O_{UV}$ are the operators 
which depend on the  SM and HS fields correspondingly. 
In  analogy with QCD one can consider, e.g., that  $O_{SM}$ is leptonic 
operator,  whereas  $O_{UV}$ is the quark operator.  
Below  $\Lambda_U$ the operator $O_{UV}$ transforms into operator of 
composite (confined) states $O_{U}$ (e.g., ``pion''): $O_{UV} \rightarrow O_{U}$
and the interaction 
becomes 
\be
C \frac{\Lambda_U^{d_{UV} - d_U}}{M^k} O_{SM}O_{U},  
\label{effopir}
\ee
where $d_{UV}$ and  $d_U$ are dimensions of operators 
$O_{UV}$ and  $O_{U}$ correspondingly. 
The key difference from the hadron case is that here 
due to scale invariance (no energy gap)  
the confined states 

\newpage
aaa
\vskip -5cm
\begin{figure}[h]
\begin{center}
\begin{minipage}{18pc}
\includegraphics[width=18pc]{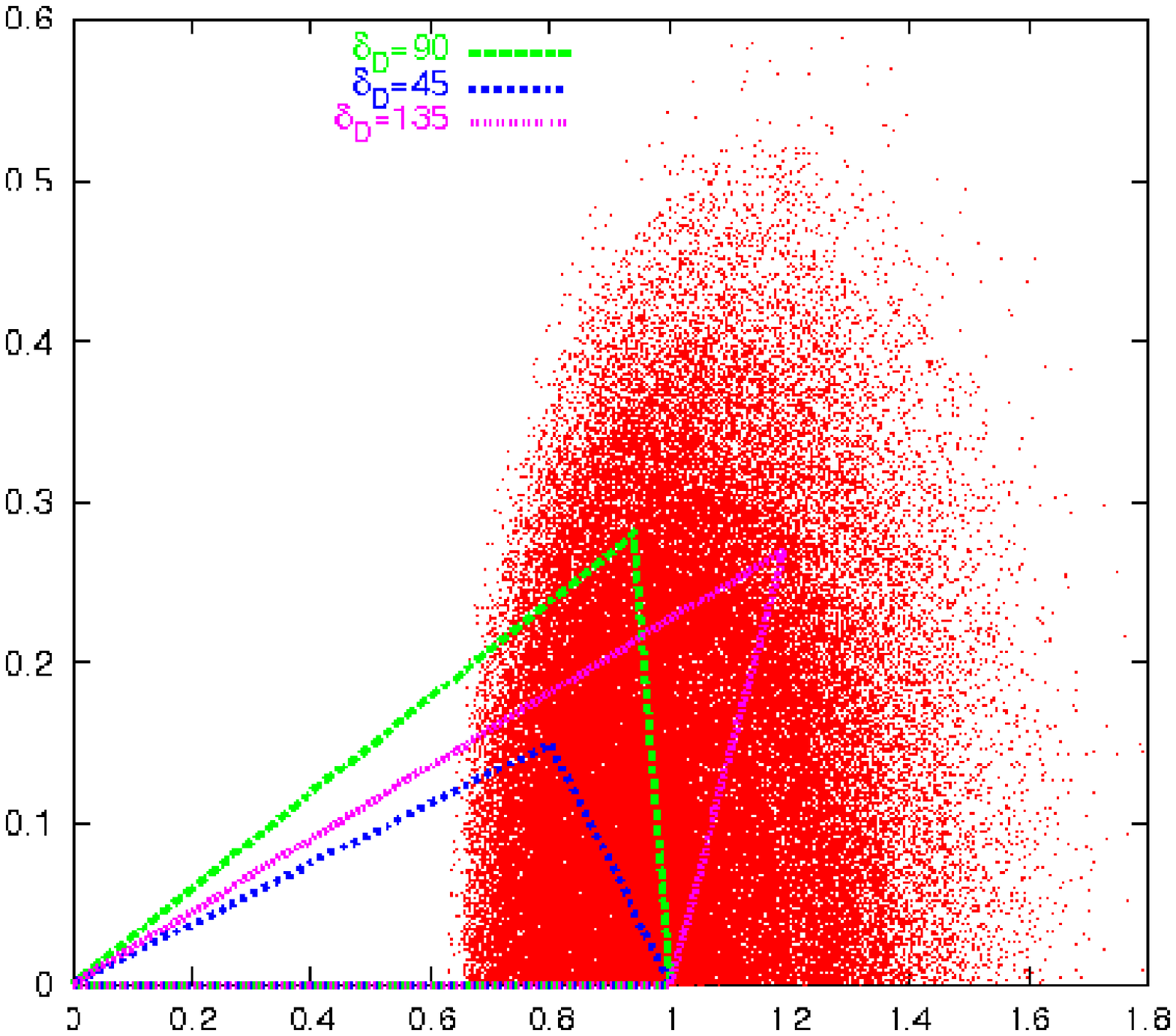}
\caption{\label{tri-a}Possible shape of the leptonic unitarity triangle now.}
\end{minipage}\hspace{1pc}%
\begin{minipage}{18pc}
\includegraphics[width=18pc]{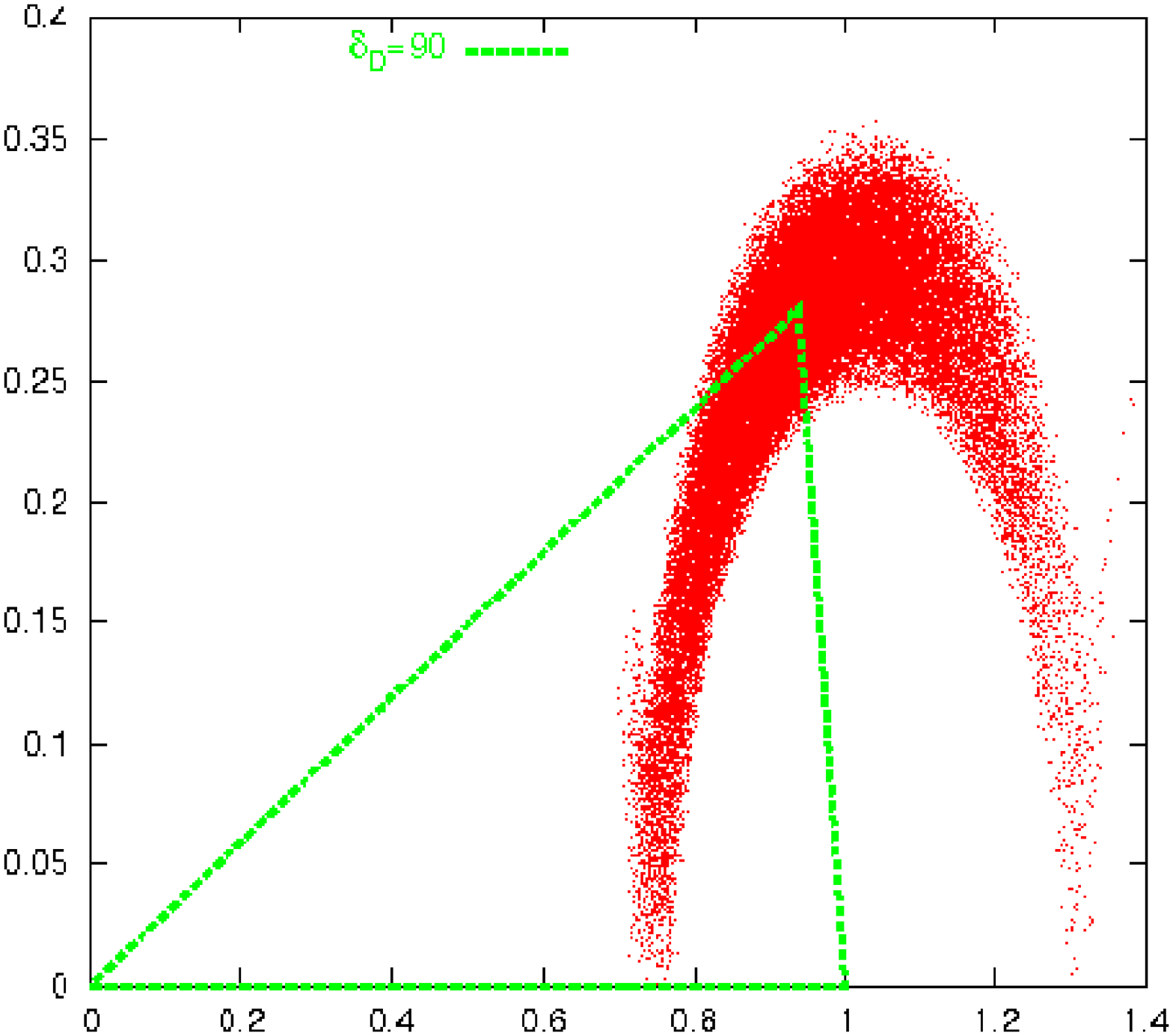}
\caption{\label{tri-b}The unitarity triangle 
in future (see text).}
\end{minipage}
\end{center}
\end{figure}

\noindent
have continuous mass spectrum
\cite{krasnikov,stepanov}.  
As a result, individual mass modes have  infinitesimal effect. 
Finite rates of production and exchange of unparticles appears 
as a consequence of integration over mass spectrum 
of composite states.

As far as applications to neutrinos are concerned,  
several processes have been considered: 
the  neutrino decays $\nu_i \rightarrow \nu_j + U$ \cite{goldberg}  
\cite{more}, scattering on electrons  
$\nu_\alpha  e \rightarrow \nu_\beta e$~\cite{more},  
and neutrino annihilation $\nu \nu \rightarrow \gamma \gamma $, 
$\nu \nu \rightarrow f f  $ \cite{annihil} via an unparticle exchange.  
The exchange of unparticles influences refraction: it modifies the matter potential
(in the case of vector operators) 
and effective neutrino mass (in the case of 
scalar operators). This, in turn, modifies 
conversion probabilities in matter \cite{concha}. 
The present data give various bounds on unparticle properties. 

\section{Future which we know}

\subsection{Reconstruction of neutrino mass and flavor spectrum} 

Clear phenomenological and experimental goal is to 
accomplish reconstruction of the neutrino 
mass and mixing spectrum. It includes measurements of 
$\theta_{13}$, the deviation of 2-3 mixing from the maximal one, 
$\delta$, absolute scale of mass, 
searches for the $\beta\beta_{0\nu}$ decay and  determination of  
nature of neutrinos, measurements of   
$m_{ee}$ and Majorana phases. 
The program has emerged more that 10 years ago. 
It is well motivated and elaborated. 
On the basis of these measurements one can reconstruct the neutrino mass 
matrix (in the flavor basis),  at least partially. 



The situation can be presented  using the leptonic unitarity triangle.  
In fig.~\ref{tri-a}  we show the possible form of 
the triangle as it follows from the existing data \cite{triangle}. 
Three $e\mu$-triangles correspond to $\sin \theta_{13} = 0.15$, 
and three different values of the CP- phase $\delta$.
The scatter-plot gives possible position of the vertex of the triangle. 
Large number of points along the 
horizontal axis corresponds to zero value of the phase 
(or $\theta_{13} = 0$). The fig.~\ref{tri-b} shows possible 
situation after the next  generation of the experiments 
(Double CHOOZ, Daya-Bay, J-PARK, NO$\nu$A)  
assuming certain  set of the results.  
Here there are no points along the horizontal axis,  
which means that non-zero value of CP-phase 
can be established, if it is not small. 
The triangle is not just an illustration,   
it may provide a method to measure $\delta$ and    
test unitarity. 

Future experimental programs are mainly based on 
the long-baseline experiments and 
the oscillograms give a global view of the situation. 
Operating and expected accelerator experiments (superbeams, beta beams, 
muon factories)  cover the energy range 
(0.5 - 30) GeV and  several baselines at $\cos \Theta_\nu < 0.3$,
that is, the peripheral regions of oscillograms with poor 
structure. This  is the origin of degeneracies of the oscillation 
parameters. 
Interesting new proposal is the low energy neutrino factory 
$E \sim few$ GeV \cite{lownuf}, which opens   
a possibility to turn the beam.

Another approach could be based on studies of the atmospheric neutrinos 
which  cover huge ranges of energies, 
$E = (0.1 - 10^4)$ GeV and base-lines, $(10 - 10^4)$ km.  
The problem here is low statistics (especially at high energies) 
and uncertainties in the original neutrino fluxes. 
The present large-scale underground and under-ice detectors 
(AMANDA, IceCube, ANTARES) have high energy thresholds, $E > (50 - 100)$ GeV,
thus missing the most interesting and structured region of the oscillograms 
at $E = (2 - 10) $ GeV.  
Both  problems can be resolved with multi-Megaton detectors 
of the TITAND type \cite{titand} with energy threshold 
below (1 - 2) GeV: high statistics will allow one to measure 
the oscillograms in a wide $E - \Theta_\nu$ range and determine 
both unknown neutrino parameters and the original fluxes 
(which can be parameterized by few quantities) simultaneously.
In such a detector  one may expect about 2000 events in 3 - 5 years,  
e.g., in the parameter space $\Delta(\cos \theta_\nu$) = 0 - 0.2 
and $\Delta E =  2 - 3$ GeV.

Measurements of the oscillograms would open a possibility to 
(i) study various oscillation effects, e.g. the  
parametric enhancement of oscillations;   
(ii) determine  the unknown neutrino parameters: the 1-3 mixing,  
mass hierarchy and  CP-phase; 
(iii) search for non-standard interactions; 
(iv) perform a tomography of the Earth with spatial resolution 
$> 100$ km.



\subsection{Searches for new physics beyond the standard scenario}


The interplay of the results of precise neutrino measurements,  
data on rare processes like 
$\mu \rightarrow e \gamma$, cosmological and astrophysical data,  
as well as results from LHC and other colliders is  expected to be very fruitful.  
Present bounds will be improved and hopefully 
signals/signatures of new physics identified. 

Neutrinos and LHC. 
Here expectations range  from complete identification  of the mechanism 
of neutrino mass generation to practically nothing. 
The first case will be realized, if, e.g., the Higgs 
triplet with a few hundred GeV  
mass and small VEV  generates neutrino mass and mixing.  
In the second one,  the outcome could be that 
some EW scale mechanisms with certain values of parameters are excluded. 
We will not be able to  detect the RH neutrinos responsible for 
the type-I seesaw mechanism. If some heavy neutral leptons 
with terascale mass are observed, they will not be immediately 
related to the light neutrino mass generation and in addition some new physics should be 
involved. The $\nu MSM-$ scenario \cite{numsm} implies yet another 
scenario of future developments.

\subsection{Neutrino astrophysics and astronomy}
Detection of neutrino bursts from galactic supernova
may have very strong impact on neutrino physics, 
astrophysics and particle physics. It can 
contribute to the determination of the neutrino parameters. 
It may shed some light on  nucleosynthesis in SN and on  
SN explosion mechanism  via the neutrino monitoring 
of the shock wave propagation. It will be an important test of 
the theory 
of neutrino propagation and  flavor  conversion. 
It is rather plausible that  we will discover 
something unexpected. 
There are good chances to measure the relic SN neutrino fluxes.

The detection of high energy neutrinos 
from astrophysical sources will be one of the major discoveries 
of this century. This will 
trigger more focused theoretical studies, 
and experimental developments. 


\section{Future which we can only imagine}

Trying to imagine future one can proceed in different ways:  
(i) ``project from the past'', e.g.  study programs of previous 
neutrino conferences;  
(ii) follow  logic of the field; 
(iii) use some historical parallels in neutrino physics and  other fields; 
(iv) imagine new neutrino sources and new detectors;  
(v) identify seeds of new developments.  


1). The breakthrough in the field can be related to developments of new 
experimental techniques: creation of new neutrino sources  and  detectors. 
This includes widely discussed beta-beams and  neutrino factories. 
Experiments with strong sources of low energy neutrinos (radioactive nuclei)
look very appealing~\cite{nostos}. 
One can imagine some particular processes at particular conditions which 
will open new  perspectives. 
One example along this line (its practical realization still should be proved)
is neutrino pair emission from metastable atoms~\cite{metastab}. 
It  looks intriguing in view of closeness of the scales of atomic transition energies 
and neutrino mass. One can expect strong enhancement of the processes 
-  superradiance due to coherence in large volume. 
The processes of photon (laser) irradiated 
neutrino pair emission from metastable atoms 
$\gamma + A_i \rightarrow \nu_i \nu_j + A_f$ and 
radiative pair emission $A_i \rightarrow \nu_i \nu_j + 
\gamma + A_f$ have been considered \cite{metastab}. 
The rates are proportional to neutrino masses and  
to Pauli blocking factor due to the presence of relic  
neutrinos. The latter can be used, in principle,  to 
detect relic neutrinos. 

On the other side, future significant progress can be due to      
development of large scintillator observatories 
and  the multi-Megaton scale water Cherenkov 
detectors with flavor (may be charge) identification and low energy thresholds.  
One can imagine new methods of light collection, 
volume detection of event, {\it etc.}. 
Further developments of the balometric techniques,  
construction of large scale array of calorimeters 
look very perspective \cite{mare}.   
The use of radioactive nuclei for neutrino detection 
with zero threshold~\cite{radnucl} opens 
some perspectives to detect relic neutrinos. 
The neutrino Moessbauer effect can be used    
to study neutrino oscillations,
measure the 1-3 mixing, determine the mass hierarchy,
search for sterile neutrinos, study gravitational redshift
of neutrinos, and even study quantum gravity effects \cite{moesappl}.
Coherent neutrino interactions can be the key feature of future 
techniques.  
One can imagine new methods of decrease  
of  background and detection of very weak signals. 
Array of km-cube size detectors of cosmic neutrinos (with KM3NET as the first step) 
is not out of discussion.


High precision of measurements will open 
new horizons to discover sub-leading effects and search for 
new physics. This in turn will trigger 
new phenomenological and theoretical studies.   

2). Neutrino structure of the Universe. 
Some work has already been done. One expects 
clumping of neutrinos depending on their masses \cite{structure}. 
That can lead to formation of neutrino halos,  and 
neutrino ``stars''. Possible new interactions (e.g.,  with 
accelerons) can lead to neutrino condensates and 
superfluidity \cite{kapusta}. 
The issue is important for the direct detection of relic neutrinos.  

The presence of relic neutrinos 
has been  established indirectly by counting the 
number of relativistic degrees of freedom
in the epoch of transition from radiation to matter dominated Universe.  
Future cosmological probes will be able to  
reconstruct structure of the Universe in the earlier epochs, 
thus resolving various  degeneracies and improving 
bounds on neutrino parameters. 
Connections neutrinos - dark energy, neutrinos - dark matter will 
be further studied. 

3). With our present advanced  knowledge of neutrino properties 
(interactions, masses and mixing) the aspect ``neutrinos as 
unique probe'' of micro and macro worlds becomes again  
important - now at a qualitatively new level.  Future experiments 
with large  fluxes and  high energy neutrinos can be used 
for further studies of the nucleon structure and 
precision measurements of the 
electroweak parameters;  NuSoNG proposal \cite{nusong}
is one step in this direction. With large-scale high statistic 
solar neutrino detectors one can have further advance in studies 
of the deep interior of the Sun, and stellar evolution  (detection of fluxes of 
N-, O-, pep-, hep - neutrinos, searches for time variations, 
correlations with solar flares, {\it etc.})
High statistics supernova neutrino detection may allow one  to monitor 
the shock wave propagation. Detection of SN bursts from other remote galaxies 
may become reality. Direct detection of the relic neutrinos will provide 
a unique  probe of the Early Universe. 


4). Toward the neutrino technologies.
Technology (applied physics) is  associated with  something
which can be copied and be of multiple use.  Some ``technologies'' 
were proposed long time ago, and  now with our accumulated 
knowledge the proposals become more realistic.  
Some examples:

-  monitoring  nuclear reactors \cite{monitor};

- oscillation and absorption tomography of the Earth; 

- study of geoneutrinos \cite{geonu}: creation of the neutrino maps of the 
Earth; 

-  use of Moessbauer neutrinos for precision measurements;

-  neutrino communication systems,
Galactic communication \cite{communic}; 

-   creation of the solar scanners to search for oil and minerals
\cite{sscanner}, {\it etc.}.

Practical realizations of at least  some of these  proposals 
look at present extremely challenging. 
At this point one can, however, recall the story of neutrinos 
themselves: in the thirties of the last century their discovery seemed to be 
impossible, 
and another  story of establishing non-zero neutrino  mass, when solution 
came not from the direct kinematic measurements but 
from the discovery of long time ``exotic'' and ``non-standard process''- 
neutrino oscillations.

\section{Conclusion}

Neutrino physics is in the transition phase. 
Significant territory is already conquered 
which can be described in terms of the standard  neutrino scenario. 
Tests of this scenario and searches 
for physics beyond it are the main objectives  
for further studies. Precision measurements and exploration 
of extreme conditions (energies, densities, 
distances) will open new horizons.

And what emerges? - Unclear implications of results for fundamental theory, 
origins of neutrino mass and mixing, the 
existence of flavor symmetries, unification,  {\it etc.}
The question what should be done to achieve  progress 
in understanding neutrino mass and mixing is already,  
and will be in future  a driving force of developments. 
LHC and other high energy experiments may clarify the situation. 

In spite of these problems we can start to think 
seriously about applied neutrino physics and  
neutrino technologies.

\end{document}